\documentclass[epj]{svjour}
\usepackage[dvips]{graphicx}
\usepackage{amsmath}

\newcommand{\tr}{\mbox{tr}}

\newcommand{\sbrace}[1]{\left( #1 \right)}

\begin{document}
\title{Hypernuclei and nuclear matter in a chiral SU(3) RMF model}
\author{Kohsuke Tsubakihara, Hideki Maekawa
         and Akira Ohnishi
}                     
%
%
\institute{Department of Physics, Faculty of Science,
           Hokkaido University, Sapporo 060-0810, Japan.\\
           \email{tsubaki@nucl.sci.hokudai.ac.jp}}
\date{Received: date / Revised version: date}
%
\abstract{
We develop a chiral SU(3) RMF model for octet baryons,
as an extension of the recently developed chiral SU(2) RMF model
with logarithmic sigma potential.
For $\Sigma$-meson coupling,
strong repulsion(SR) and weak repulsion(WR)
cases are examined in existing atomic shift data of $\Sigma^-$.
In both of these cases, we need an attractive pocket of a few MeV depth
around nuclear surface.
} 
\titlerunning{
Hypernuclei and nuclear matter in a chiral SU(3) RMF model
}
\authorrunning{Tsubakihara {\em et al.},}
\maketitle
\section{Introduction}
\label{intro}
In constructing the dense matter equation of state (EOS),
it is strongly desired to respect
both of hypernuclear physics information and chiral symmetry.
Strangeness is expected to play a decisive role
and the partial restoration of chiral symmetry
would modify the hadron properties 
in dense matter.
One of the promising approaches is to apply chiral symmetric relativistic
mean field (RMF) models~\cite{Boguta,Instability,TooStiff,Glueball,Papazoglou1998,Tsubaki2006,Zschiesche:2006zj}.

We have recently developed
a chiral SU(2) symmetric RMF model~\cite{Tsubaki2006}
with logarithmic sigma potential in the form of $-\log\sigma$,
which is derived in the strong coupling limit (SCL)
of the lattice QCD~\cite{SCL}.
In this model, the energy density in vacuum at zero temperature is evaluated 
in the mean field approximation as,
\begin{align}
&  U_\sigma
  =  - a\,\log(\det MM^\dagger)
      + b\,\tr(MM^\dagger)
      + c_\sigma\sigma\nonumber\\
&~~~\sim
 - 2a\,f_\mathrm{SCL}(\frac{\sigma}{f_\pi})
      + \frac{1}{2}m_\sigma^2\sigma^2\ ,
\label{SU2_EI}\\
    & f_\mathrm{SCL}(x)
        = \log(1-x)+x+\frac{x^2}{2}\ ,
    \quad
    a  = \frac{f_\pi^2}{4}\,(m_\sigma^2-m_\pi^2) \ ,
        \nonumber
\end{align}
where $M$ denotes the SU(2) meson matrix,
$M=(\sigma+i{\boldsymbol{\pi}}\cdot{\boldsymbol{\tau}})/\sqrt{2}$.
In the second line of Eq.~(\ref{SU2_EI}),
$\sigma$ field is replaced with its fluctuation around the vacuum expectation
value, $\sigma \to f_\pi - \sigma$.
In this SCL model~\cite{Tsubaki2006},
we can describe bulk properties of finite nuclei,
we have neither the chiral collapse at low densities~\cite{Boguta}
nor instability at large $\sigma$~\cite{Instability},
and the nuclear matter EOS is not very stiff~\cite{TooStiff}.
Compared to previously proposed~chiral RMF models~\cite{Glueball,Papazoglou1998}
and a more recently proposed one~\cite{Zschiesche:2006zj},
this model has an advantage that the vacuum energy density
is derived based on QCD.

It is straightforward to extend this chiral SU(2) RMF model to
an SU(3) version which contains strangeness degrees of freedom.
We expect that this extension enables us
to get detailed information on $\Lambda$, $\Sigma$ and $\Xi$ hypernuclei.

In this paper, we determine the hyperon-meson coupling constants
in this chiral SU$_{f}$(3) RMF model by fitting existing data.
We show that we can reproduce the separation energies of
single $\Lambda$ hypernuclei ($S_\Lambda$)~\cite{LambdaData}
and the $\Lambda\Lambda$ bond energy ($\Delta B_{\Lambda\Lambda}$)
in $^6_{\Lambda\Lambda}$He~\cite{Nagara}
by choosing the coupling constants appropriately
in a reasonable parameter range.
The EOS of symmetric matter is found to be softened
by the scalar meson with hidden strangeness, $\zeta=\bar{s}s$,
which couples with $\sigma$ through the determinant interaction.
We also discuss the strength of repulsion in nuclear medium
and attraction around nuclear surface in $\Sigma^-$-nucleus potential
by comparing the calculated results with
$\Sigma^-$ atomic shift data~\cite{SigmaData}.

\section{Chiral SU(3) RMF model}
\label{sec:2}

In extending the chiral SU(2) RMF model to SU(3),
it is necessary to include mesons with hidden strangeness ($\bar{s}s$)
such as $\zeta$ and $\phi$ in addition to $\sigma$, $\omega$ and $\rho$.
The chiral SU(2) RMF model~\cite{Tsubaki2006}
tells us the form of chiral potential of $\sigma$ and $\zeta$
by a simple extension written as,
\begin{align}
  U_{\sigma\zeta}
  = & - a\,\log(\det MM^\dagger)
      + b\,\tr(MM^\dagger)\nonumber\\
    & + c_\sigma\sigma + c_\zeta\zeta 
      + d\,(\det M + \det M^\dagger),
    \label{chiralSU3int}
\end{align}
where the last term in rhs is introduced to take care of
the U$_A$(1) anomaly.
When the chiral symmetry
is spontaneously broken
and meson mass terms are generated,
this effective interaction is written as,
\begin{align}
  U_{\sigma\zeta}
  = & - a\left[2f_{\mbox{\scriptsize SCL}}(\frac{\sigma}{f_\pi}) 
                 + f_{\mbox{\scriptsize SCL}}(\frac{\zeta}{f'_\zeta}) 
		 \right]
                 \nonumber\\
    & + \frac{1}{2}m_\sigma^2\sigma^2 + \frac{1}{2}m_\zeta^2\sigma^2
      + \xi_{\sigma\zeta}\sigma\zeta\ ,
    \label{chiralSU3int_mass}
\end{align}
where $f'_\zeta = f_\zeta + m_s$ 
and $m_s$ is related to the strange quark mass.
We have six parameters in this interaction
($a$, $b$, $c_\sigma$, $c_\zeta$, $d$ and $m_s$),
and five out of six are fixed by fitting experimental masses
of $\pi$, $K$ and $\zeta$,
and vacuum expectation values of $\sigma$ and $\zeta$.
There remains only one parameter, $m_\sigma$.
With this scalar meson effective interaction,
the RMF Lagrangian is given as,
\begin{align}
\mathcal{L} =
 & \mathcal{L}_\mathrm{Free}
  (\psi_i, \bar{\psi}_i, \sigma, \zeta, \omega, \rho, \phi)
  +\mathcal{L}_\mathrm{EM}
  - U_{\sigma\zeta}
  + \frac{c_\omega}{4}\omega^4 \nonumber\\
 + & \sum_i\bar\psi_i \left[
       g_{\sigma i}\sigma + g_{\zeta i}\zeta
      - \gamma_\mu(
      g_{\omega i}\,{\omega}^\mu
      + g_{\rho i}\,{\rho}^\mu
      + g_{\phi i}\,{\phi}^\mu
          )
      \right] \psi_i
      \ ,
\label{SU3Lag}
\end{align}
where the $\omega^4$ term is phenomenologically introduced
to simulate the high density behavior of the vector self-energy
in the RBHF theory as in Ref.~\cite{ST94}.

In determining hyperon-vector meson couplings,
we start from the SU$_f$(3) symmetric interaction,
\begin{align}
\mathcal{L}_\mathrm{BM}
= \sqrt{2}\{ g_s\,\tr\sbrace{M} \tr\sbrace{\bar{B}B}
                   & + g_1\,\tr\sbrace{\bar{B}MB} \nonumber\\
		   & + g_2\,\tr\sbrace{\bar{B}BM}\}\ .
\label{SU3coupling}
\end{align}
Following the Okubo-Zweig-Iizuka (OZI) rule~\cite{OZI},
we assume that nucleons do not
couple with $\bar ss$ mesons ($\zeta$ and $\phi$).
Then
there are two independent parameters, $g_{\omega N}$ and $g_{\rho N}$,
and hyperon-vector meson coupling constants are found
to be represented by $g_{\omega N}$ and $g_{\rho N}$ as follows,
\begin{eqnarray}
&&
g_{\omega\Lambda} = \frac{5}{6}g_{\omega N} - \frac{1}{2}g_{\rho N},\;\;
g_{\phi\Lambda} = \frac{\sqrt{2}}{3}\sbrace{g_{\omega N} + 3g_{\rho N}}\ ,
\label{LambdaVM}
\\
&&
g_{\omega\Sigma}
= g_{\rho\Sigma}
= \frac{g_{\phi\Xi}}{\sqrt{2}}
= \frac12(g_{\omega N} + g_{\rho N})\ ,
\label{SigmaVM}
\\
&&
g_{\omega\Xi} = g_{\rho\Xi}
= \frac{g_{\phi\Sigma}}{\sqrt{2}}
=\frac12(g_{\omega N} - g_{\rho N})\ .
\label{XiVM}
\end{eqnarray}
In the later discussion,
we try to keep the above relations
as far as possible.

In the scalar and pseudo scalar sector,
it is necessary to include negative parity baryons
or we only have D-type
when the chiral SU(3)
symmetry
is required~\cite{Papazoglou1998,SU3MB}.
This problem is out of the scope of this proceedings,
and hyperon-scalar meson coupling constants are regarded as parameters.
When the $\Lambda$-scalar meson couplings are obtained
and SU$_f$(3) symmetry works also for scalar couplings,
we can evaluate the $\Xi$-scalar couplings as,
\begin{equation}
g_{\sigma\Xi} = \frac{2}{3}g_{\sigma N} - \frac{\sqrt{2}}{2}g_{\zeta\Lambda},\;\;
g_{\zeta\Xi}  = \frac{1}{3}g_{\sigma N} + \frac{\sqrt{2}}{2}g_{\zeta\Lambda}.
\label{XiSM}
\end{equation}

\section{Nuclear matter and hypernuclei}
\subsection{Normal nuclei and nuclear matter}

\begin{figure}[t!]
\centering
\resizebox{0.48\textwidth}{!}{%
  \includegraphics{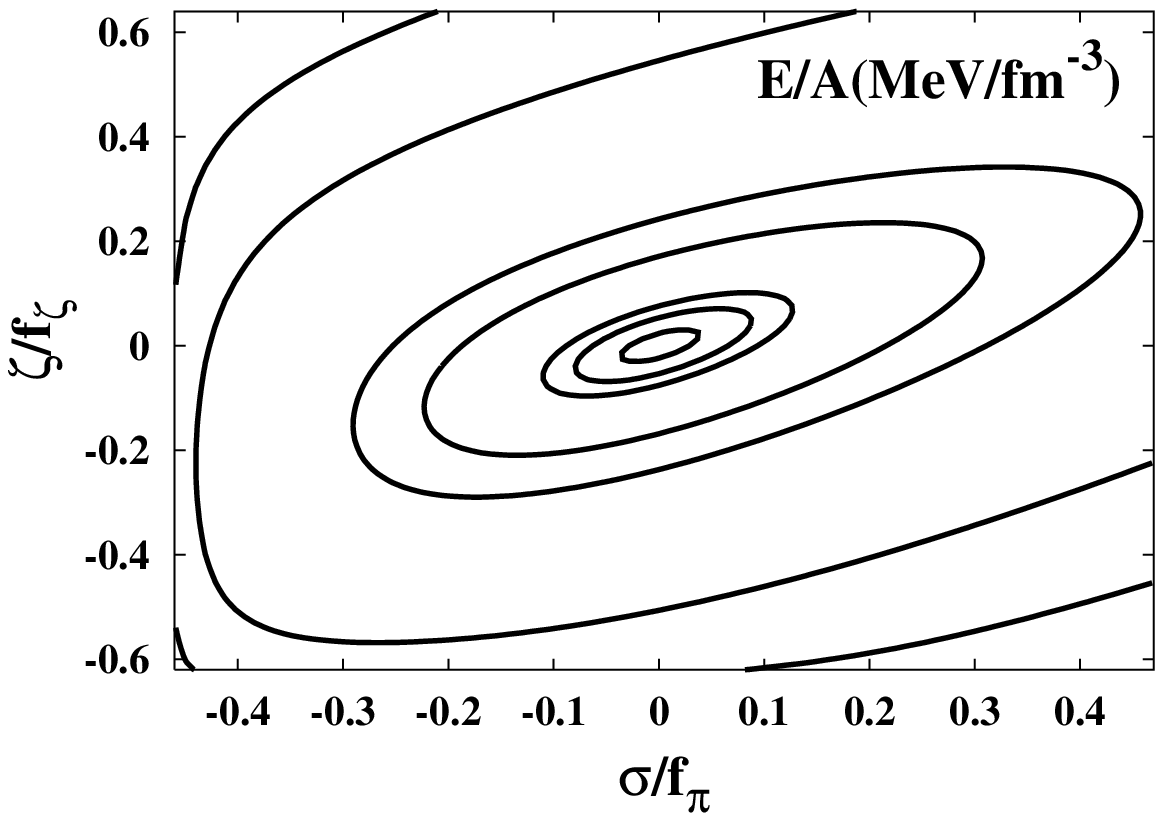}
}
\resizebox{0.48\textwidth}{!}{%
  \includegraphics{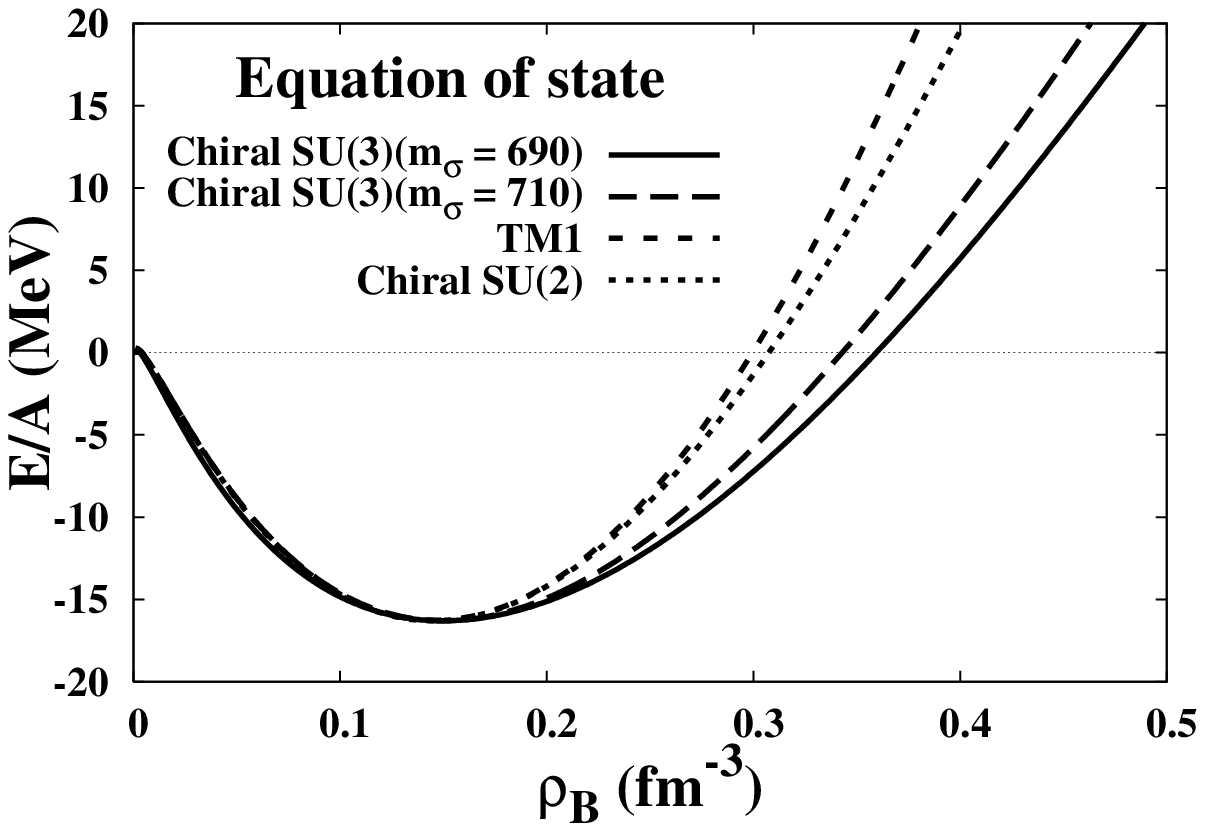}
}
\caption{Energy surface and EOS in chiral SU(3) model.}
\label{fig:1}       
\end{figure}

In the present chiral RMF model,
bulk properties of normal nuclei are well described,
and these results are reported elsewhere.
The strangeness degrees of freedom are found to soften the nuclear matter EOS,
and thus have effects also on normal nuclei.
The interaction in Eq.~(\ref{chiralSU3int_mass}) contains
the $\sigma\zeta$ mixing term,
which gives rise to a correlation in $\sigma$ and $\zeta$
along the softest valley in the vacuum energy surface 
as shown in the upper panel of Fig. \ref{fig:1}.
Since the matter can evolve along this valley as the density increases,
EOS is softened than in the chiral SU(2) RMF model~\cite{Tsubaki2006},
in which there is no $\zeta$ degree of freedom.
The incompressibility is found to be $K \sim 220$ MeV
when we fit
the bulk properties of normal nuclei and nuclear matter saturation point,
as shown in the lower panel of Fig. \ref{fig:1}.

\subsection{$\Lambda$ hypernuclei}

\begin{figure}[t!]
\resizebox{0.48\textwidth}{!}{%
  \includegraphics{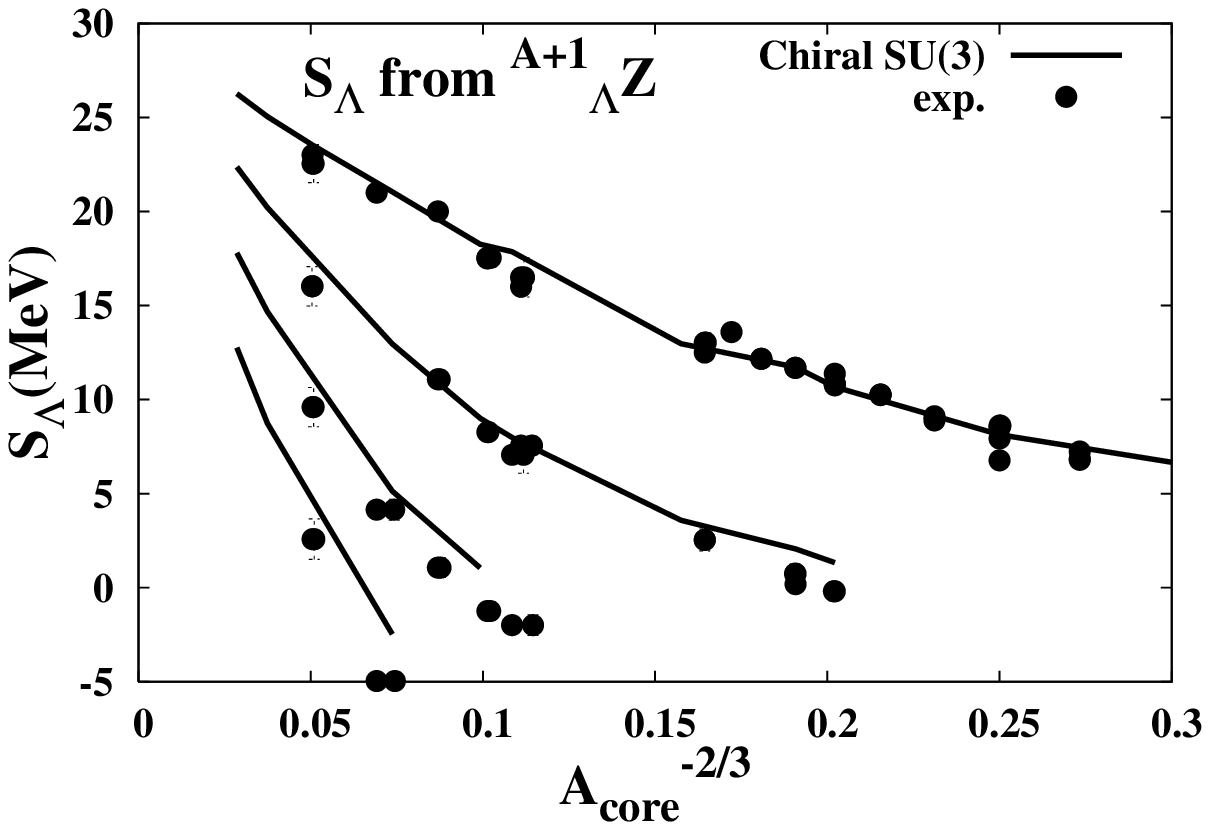}
}
\resizebox{0.48\textwidth}{!}{%
  \includegraphics{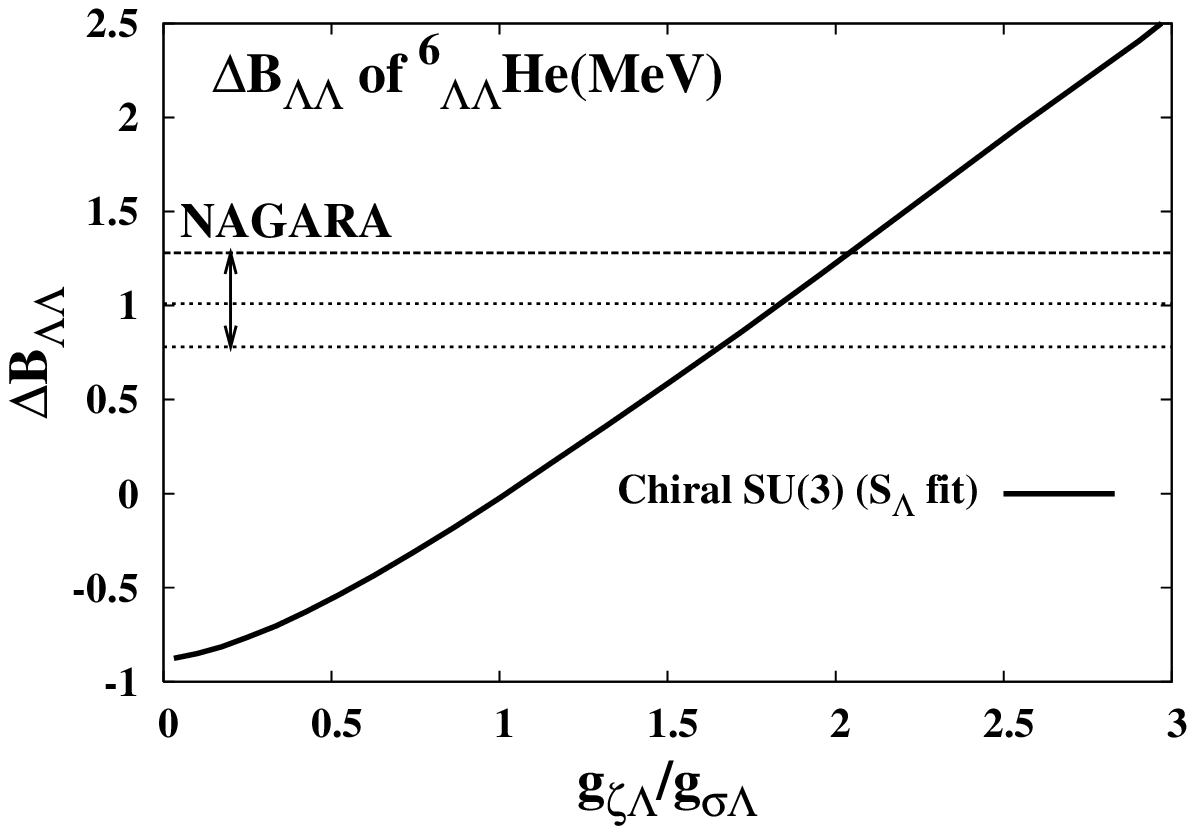}
}
\caption{$\Lambda$ separation energy and
$\Delta B_{\Lambda\Lambda}$ of $^6_{\Lambda\Lambda}$He. }
\label{fig:2} 
\end{figure}

Next we study $\Lambda$ hypernuclei with this chiral SU(3) RMF Lagrangian.
There appear four additional parameters, 
$g_{\sigma\Lambda}$, $g_{\zeta\Lambda}$, $g_{\omega\Lambda}$
and $g_{\phi\Lambda}$.
We fix the vector coupling constants, $g_{\omega\Lambda}$ and $g_{\phi\Lambda}$
by using the SU(3) symmetry relation in Eq. (\ref{LambdaVM}).
Two remaining parameters are determined
by fitting $S_\Lambda$ and $\Delta B_{\Lambda\Lambda}$ data.
As shown in the upper panel of Fig.~\ref{fig:2},
we can explain $S_\Lambda$ nicely in a wide mass region
by giving the $\Lambda$ potential depth around 30 MeV,
which is represented by a linear combination of
$g_{\sigma\Lambda}$ and $g_{\zeta\Lambda}$.
By fitting $\Delta B_{\Lambda\Lambda}$ in ${}^6_{\Lambda\Lambda}\mathrm{He}$
simultaneously with $S_\Lambda$,
both of $g_{\sigma\Lambda}$ and $g_{\zeta\Lambda}$ are determined
as shown in Fig. \ref{fig:2}.

\subsection{$\Sigma$ hyper atom}
\label{Sec:3}

Recent analyses of quasi-free $\Sigma^-$ production spectra
~\cite{SigmaQF,MTO07}
suggest that $\Sigma^-$-nucleus potential should be repulsive in nuclear medium.
On the other hand, $\Sigma^-$-nucleus potential needs to possess a few MeV 
attractive pocket around nuclear surface
to explain $\Sigma^-$ atomic shift data~\cite{B94,M95}.
Here we would like to extract $\Sigma$-meson coupling constants
which explain $\Sigma^-$ atomic shifts.
In the present RMF model, we have four additional parameters for $\Sigma$,
$g_{\sigma\Sigma}$, $g_{\zeta\Sigma}$, $g_{\omega\Sigma}$ and $g_{\rho\Sigma}$. 
First we set $g_{\omega\Sigma}$,
which determines the strength of repulsion in nuclear medium.
We have examined two cases.
(i) Strong Repulsion (SR) case:
From the flavor SU(3) symmetry and OZI rule,
$g_{\omega\Sigma}$ is given as
$g_{\omega\Sigma} = (g_{\omega N} + g_{\rho N})/2 \sim 2 g_{\omega N}/3$.
(ii) Weak Repulsion (WR) case:
$g_{\omega\Sigma}\sim g_{\omega N}/3$ which is also adopted
in Ref. \cite{M95}.
Secondly, scalar meson couplings ($g_{\sigma\Sigma}$ and $g_{\zeta\Sigma}$),
which determine the attractive pocket depth around nuclear surface,
are chosen so as to reproduce atomic shifts of symmetric $N=Z$ core nuclei
(O, Si, S).
Finally, $g_{\rho\Sigma}$ is adjusted to get a correct atomic shift in Pb.

\begin{figure}[t!]
\centering
\resizebox{0.48\textwidth}{!}{%
  \includegraphics{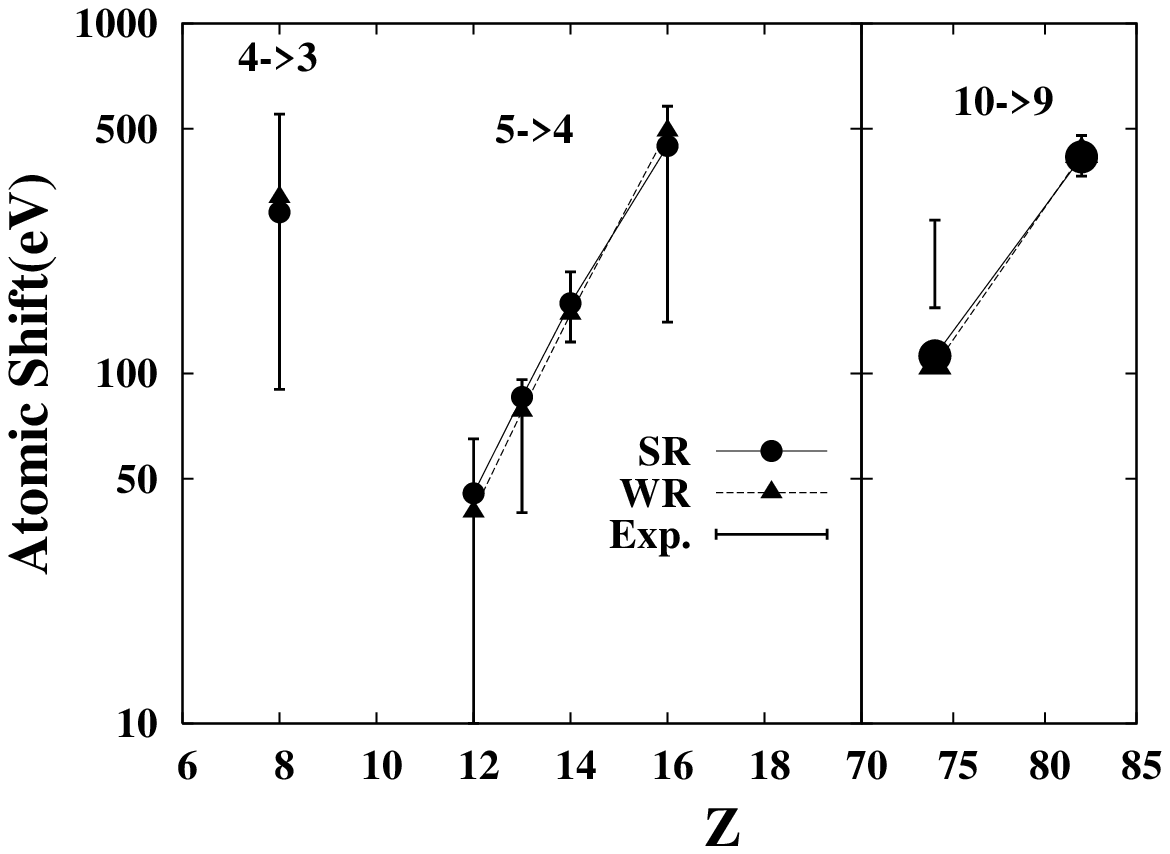}
}
\resizebox{0.48\textwidth}{!}{%
  \includegraphics{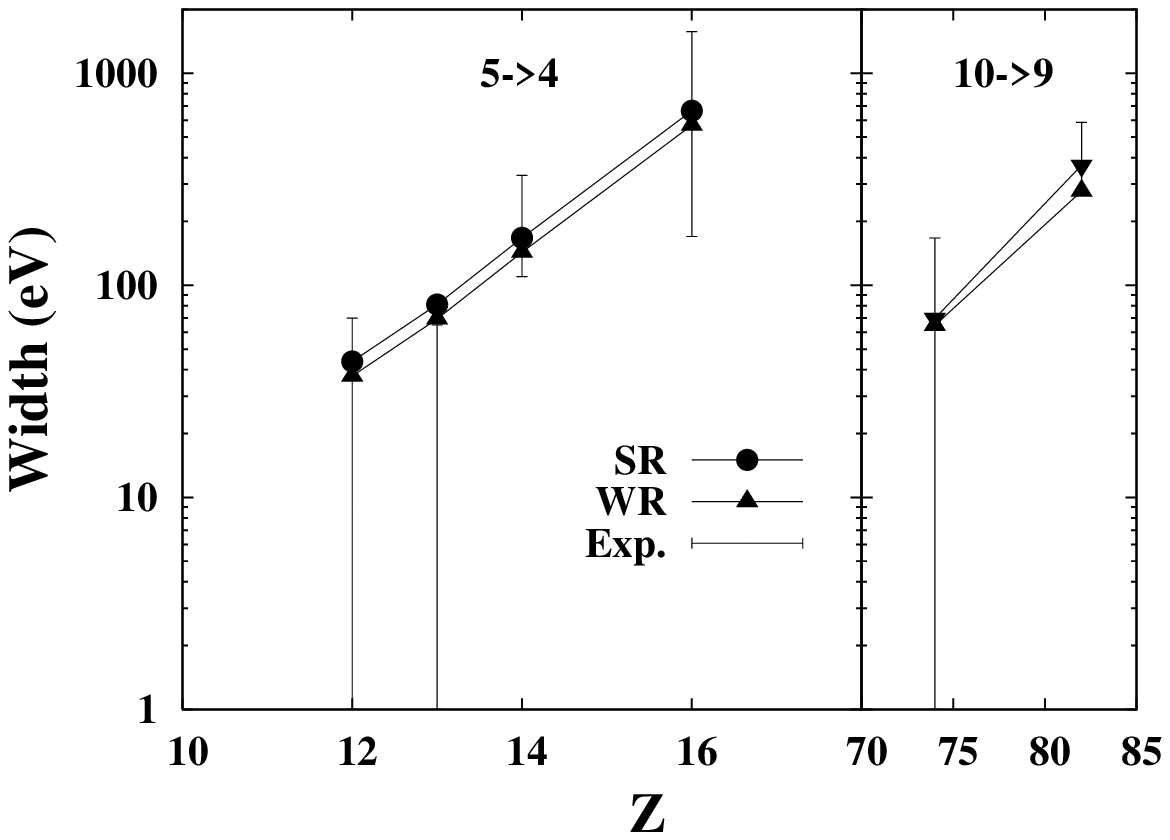}
}
\caption{Atomic shift and conversion width of $\Sigma^-$.}
\label{fig:3}
\end{figure}

In Fig. \ref{fig:3},
we show calculated atomic shifts and conversion widths
of O, Mg, Al, Si, S, W and Pb
for $n=4\rightarrow3$(O),
$n=5\rightarrow4$(Mg, Al, Si and S)
and $n=10\rightarrow9$(W and Pb) transitions.
Atomic shift results are in good agreement except for W
and the total $\chi^2 /$ $\mbox{dof}$ is around $1.3$.
The conversion width is calculated as 
the expectation value of $\mbox{Im}V_{\mbox{opt}} = t \rho_p$.
Imaginary parts are found to be $- 15 \sim - 20$ MeV.

\begin{figure}[t!]
\centering
\resizebox{0.48\textwidth}{!}{%
  \includegraphics{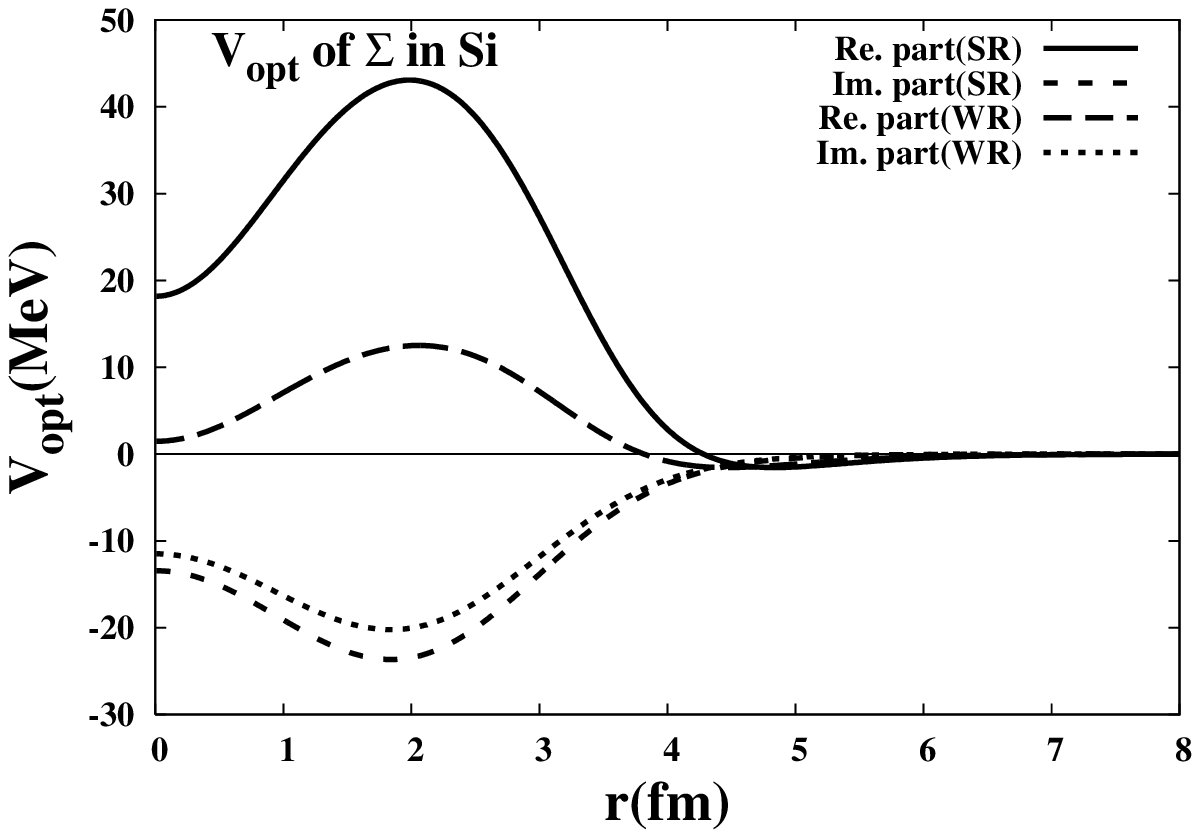}
}
\resizebox{0.48\textwidth}{!}{%
  \includegraphics{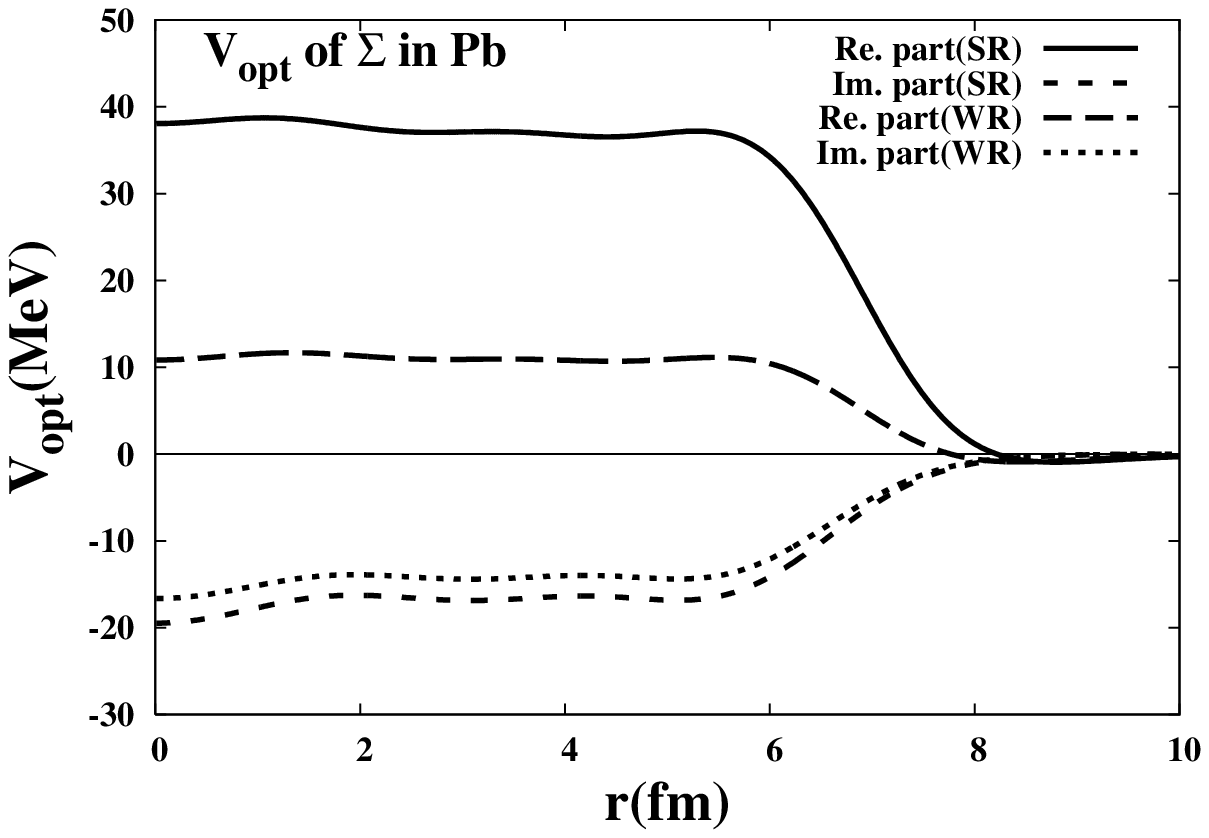}
}
\caption{Re.\ part and Im.\ parts of optical potentials of $\Sigma^-$
in SR and WR cases. 
}
\label{fig:4}       
\end{figure}

\section{Summary and conclusion}

We have developed a chiral SU(3) relativistic mean field (RMF) model
with a logarithmic chiral potential
for $\sigma$ and $\zeta(=\bar{s}s)$ mesons
derived in the strong coupling limit of lattice QCD~\cite{SCL},
as an extension of the chiral SU(2) RMF model~\cite{Tsubaki2006}.
The chiral symmetry and the mass generation
by the spontaneous chiral symmetry breaking
give severe constraints on parameters.
After fitting several meson masses and vacuum expectation values,
$m_\sigma$ is left unfixed in this chiral potential.
Nucleon parameters
($N$-meson coupling constants, $m_\sigma$
and the coefficient of $\omega^4$ term)
are determined to reproduce
the vacuum nucleon mass,
the nuclear matter saturation point, and
bulk properties (binding energies and charge rms radii)
of normal nuclei from C to Pb isotopes.
$\Lambda$-meson coupling constants are 
determined by fitting hypernuclear data
($\Lambda$ separation energies $S_\Lambda$
 and $\Lambda\Lambda$ bond energy $B_{\Lambda\Lambda}$) 
under the constraints of SU$_f$(3) symmetry for vector couplings.

By fitting the $\Sigma^-$ atomic shifts,
we find that the attractive pocket in the $\Sigma$-nucleus potential
around the nuclear surface should have a few MeV depth.
The conversion widths of $\Sigma^-$ atom are well described
with the imaginary part of the optical potential
in the form of $\mbox{Im}V_{\mbox{opt}}=t\rho_p$,
and the strengths are found to be $-15 \sim -20$ MeV.
These results are consistent with the previous RMF analysis~\cite{M95}.

We have tried to keep the SU$_f$(3) relations
in the baryon-vector meson coupling constants as far as possible,
and these relations seem to work well for $\Lambda$ hypernuclei.
However, we have to break the SU$_f$(3) relation for $g_{\rho\Sigma}$
to reproduce atomic shift data at $N > Z$.
It is suggested that the short-range repulsion is strong in $\Sigma N$
interaction due to the Pauli blocking between quarks.
Therefore, the present result may indicate
that we cannot describe $\Sigma^-$-nucleus potential properly
in the chiral SU(3) RMF,
which should be applicable to hadronic interactions,
and that it is necessary to include the short-range repulsion
from quarks.
While we have this conceptual problem, 
we have now a chiral SU(3) RMF model,
which can describe nuclear matter, finite normal nuclei,
single and double $\Lambda$ hypernuclei and $\Sigma^-$ atom.

It is desired to check the consistency between the present results
and quasi-free spectrum analyses.
We have investigated $\Sigma^-$ quasi-free spectrum with
DWIA+Local Optimized Fermi Average t-matrix~\cite{MTO07}.
With this method, it would be possible to judge
whether $\Sigma^-$ repulsion should be strong or relatively weak.
It is also interesting to investigate $\Xi$ hypernuclei and hyperatoms.
If the SU$_f$(3) relations in Eqs.~(\ref{XiVM}) and (\ref{XiSM})
approximately hold in $\Xi$-meson couplings,
we have smaller 
ambiguities in the $\Xi$-nucleus potential.
Predictions along this line are in progress.

%
%
\section*{Acknowledgment}
This work is supported in part by the Ministry of Education,
Science, Sports and Culture, Grant-in-Aid for Scientific Research
under the grant numbers,
    15540243		
and 1707005.		

\end{document}